\documentclass[reqno,11pt]{amsart}
\usepackage{amsmath,amssymb,tabularx,setspace,color,multirow,graphics,graphicx}
\newtheorem{theorem}{Theorem}
\usepackage[margin=3.5cm]{geometry}

\newtheorem{definition}{Definition}
\newtheorem{remark}{Remark}

\makeatletter
\renewcommand\section{\@startsection {section}{1}{\z@}%
                                   {-3.5ex \@plus -1ex \@minus -.2ex}%
                                   {2.3ex \@plus.2ex}%
                                   {\normalfont\large\bfseries}}
\makeatother

\newtheorem{Corollary}{Corollary}

\newtheorem{Theorem}{Theorem}

\newtheorem{Example}{Example}

\begin{document}
\doublespace
\vspace{-0.3in}
\title[]{Non-parametric estimation of cumulative (residual) extropy }
\author[]%
{   S\lowercase{udheesh} K. K\lowercase{attumannil\textsuperscript{a} and } S\lowercase{reedevi} E. P.\textsuperscript{\lowercase{b}}\\
 \lowercase{\textsuperscript{a}}I\lowercase{ndian} S\lowercase{tatistical} I\lowercase{nstitute},
  C\lowercase{hennai}, I\lowercase{ndia},\\ \lowercase{\textsuperscript{b}}SNGS C\lowercase{ollege}, P\lowercase{attambi}, I\lowercase{ndia}.}
\thanks{{$^{\dag}$}{Corresponding E-mail: \tt sreedeviep@gmail.com}}
\maketitle
\vspace{-0.2in}

\begin{abstract}Extropy and its properties are explored to quantify the uncertainty. In this paper, we obtain  alternative expressions for cumulative residual extropy and negative cumulative  extropy. We obtain  simple estimators of  cumulative (residual)  extropy.   Asymptotic properties of the proposed estimators are studied.  We also present new estimators of cumulative (residual)  extropy  when the data is right censored.  The finite sample performance of the estimators is evaluated through  Monte Carlo simulation studies.  We use the proposed estimators to analyse different real data sets. Finally, we  obtain the relationship  between different  dynamic and weighted extropy measures  and reliability concepts,   which leads to  several open problems associated  with these measures.
\end{abstract}
\vspace{-0.1in}
\section{Introduction}
The uncertainty associated with a random variable can be evaluated using information measures. The widely used  measure of information theory is entropy. 
For an absolutely continuous and non-negative random variable $X$, the differential entropy proposed by Shannon (1948)  is given by
\begin{equation*}
H(X)=-E(\text{log}f(X))= \int_{0}^{\infty}f(x)\text{log}(f(x))dx,
\end{equation*}
where $f(x)$ is the probability density function of $x$, provided above integral exists. Later, various measures of entropy are defined in the literature, each one suitable for specific situations. The widely used versions of entropy measures involved  cumulative residual entropy (Rao et al., 2004), cumulative entropy (Di Crescenzo and Longobardi, 2009) and the corresponding weighed measures by Mirali et al. (2016) and Mirali and Baratpour (2017). Recently an alternative measure of uncertainty called extropy is proposed by
Lad et al. (2015) as the complementary dual of entropy as following.  For non-negative random variable  $X$, extropy is defined as
$$J(X)=-\frac{1}{2}\int_{0}^{\infty}f^2(x)dx.$$
For more properties of $J(X)$ see Lad et al. (2015). Recently, several researchers studied various forms of extropy and proposed some new measures. Extropy of order statistics and record values is studied by Qiu (2017) and Jose and Sathar (2019) among others. Qiu and Jia (2018a) studied  the residual extropy of order statistics.  Qiu and Jia (2018b) developed goodness of fit of uniform distribution using extropy. Qiu et al. (2019) studied the properties of extropy for mixed systems. Jahanshahi et al. (2020) and Tahmasebi and Toomaj (2020) studied cumulative residual extropy and negative cumulative extropy in detail.  Recently, Balakrishnan et al. (2020), Banal and Gupta (2021) and Sathar and Nair (2021a, 2021b, 2021c) studied different weighted versions of extropy.


The rest of the article is organised as follows. In Section 2, we provide an alternative expression for cumulative residual extropy and then obtained an estimator of the same.  We also propose a new estimator  of cumulative residual extropy  when the data contain right censored observations.  In Section 3, we obtain  simple non-parametric estimators of the negative cumulative extropy for complete and right censored data.  We prove  that  proposed   estimators are consistent and  asymptotically normally distributed. In section 4, we report the results of the  Monte Carlo simulation study  conducted to evaluate the finite sample performance of the proposed estimators.  Analysis of several real data sets using the proposed estimators are given in Section 5.  In Section 6, we obtain the relationship between different extropy measures and reliability concepts which leads  to several open   problems in this area.

\section{Cumulative residual extropy}\vspace{-.1in}
Let $X$  be a non-negative continuous random variable with density and distribution function $f$ and $F$, respectively. Define survival function of $X$ at $x$ as $\bar{F}(x)=1-F(x).$   Assume that $X$ has finite mean $\mu$.
The cumulative residual extropy is given by (Jahanshahi et al., 2020)
\begin{equation}\label{creext}
   \mathcal{CRE}=-\frac{1}{2}\int_{0}^{\infty}\bar{F}^2(x)dx.
\end{equation}
Next we discuss the estimation of $\mathcal{CRE}$  under complete and censored situations.
\subsection{Uncensored case}
\vspace{-.1in}
 Let $X_1$ and $X_2$ be two independent  random variables having  distribution function $F$. Then the survival function of  $\min(X_1,X_2)$ is given by $\bar{F}^2(x)$.
For a positive random variable $X$ we have
$$E(X)=\int_{0}^{\infty}xf(x)dx=\int_{0}^{\infty}\bar{F}(x)dx.$$
Hence we can represent $ \mathcal{CRE}$ as\vspace{-.1in}
\begin{equation}\label{adef}
  \mathcal{CRE}=-\frac{1}{2}E(\min(X_1,X_2)).
\end{equation}
An estimator of $\mathcal{CRE}$ based on  U-statistics is given by
\begin{equation*}
  T_1=\frac{-1}{n(n-1)}\sum_{i=1}^{n-1}\sum_{j=i+1}^{n}\min(X_i,X_j).
\end{equation*}Clearly $T_1$ is an unbiased estimator of $ \mathcal{CRE}$. Using order statistics we can express above estimator as
\begin{equation}\label{estcre}
  T_1=\frac{-1}{n(n-1)}\sum_{i=1}^{n}(n-i)X_{(i)},
\end{equation}where $X_{(i)}$ denote the $i$-th order statistics based on a random sample $X_1,\ldots,X_n$ from $F$.
Next we study the asymptotic properties of $T_1$ using the limit theorems of U-statistics. Since $T_1$ is a U statistic, it is consistent estimator of $ \mathcal{CRE}$ (Lehmann, 1951).
 \vspace{-0.1in}
\begin{theorem} As $n\rightarrow \infty$,  $\sqrt{n}( T_1-\mathcal{CRE})$  converges in distribution to  Gaussian random variable  with mean zero and variance $4{\sigma_1^2}$ where \vspace{-0.1in}
\begin{equation}\label{eq8} \vspace{-0.14in}
 \sigma_1^2 = V\Big(X\bar{F}(X)+\int_{0}^{X}ydF(y)\Big).
\end{equation}
\end{theorem}
\noindent{\bf Proof: } By central limit theorem of U-statistics, we have the asymptotic normality. The asymptotic variance is $4\sigma_1^2$, where $\sigma_1^2$ is given by (Lee, 2019)
\begin{equation*}
 \sigma_1^2= Var\left(E(\min(X_1,X_2)|X_1)\right).
\end{equation*} Consider
\begin{eqnarray*}
  E(\min(X_1,X_2)|X_1=x)&=&E(xI(x< X_2)+X_2I(X_2\le x ))\\&=&x\bar{F}(x)+\int_{0}^{x}ydF(y).
\end{eqnarray*}
Hence we have the variance expression specified as in the theorem.\\
\subsection{Right censored case}
In this section, we obtain a simple estimator of $\mathcal{CRE}$ for right censored data. Suppose we have randomly right-censored observations where the censoring times are independent of the lifetimes. Let  $C$ be the censoring random variable with survival function $K$.  We  are interested to find an estimator of the cumulative residual extropy  based on $n$ independent and identical observations $\{(Y_i,\delta_i),1\leq i\leq n\}$, where $Y_i=\min(X_i,C_i)$ and $\delta_i=I(X_\leq C_i)$, is the censoring indicator.  A U-statistic defined for censored data is given by (Datta et al., 2010)
\begin{equation}\label{eq4.3}
 {T}_{1c}=-\frac{1}{n(n-1)}\sum_{i=1}^{n}\sum_{j<i;j=1}^{n}\frac{\min(Y_i,Y_{j})\delta_i\delta_j}{\widehat{K}(Y_i-)\widehat{K}(Y_j-)},
\end{equation}
where $\widehat{K}(.)$ is the Kaplan-Meier estimator of $K(.)$.  
\noindent Now we study the asymptotic properties of $  {T}_{1c}$. In the next theorem, we state the consistency of  ${T}_{1c}$. The following results can be proved in similar lines as Theorem 2 of Sudheesh et al. (2021).
\begin{theorem}As $ n \rightarrow \infty $, $ {T}_{1c} $ converges in probability  to $\mathcal{CRE}$.
\end{theorem}

\par To derive the  asymptotic distribution of  $ {T}_{1c}$, define \begin{small}$N_i^c(t)=I(Y_i\leq t, \delta_i=0)$\end{small} as the counting process corresponding to censoring for the $i$-{th} subject  and  $R_i(u)=I(Y_i\geq u)$. Let $\lambda_c(t)$ be the hazard rate of the censoring  variable $C$. The martingale associated with the counting process $N_i^c(t)$ is given by
\begin{equation*}
M_i^c(t)=N_i^c(t)-\int_{0}^{t} R_i(u) \lambda_c(u) du.
\end{equation*}
Let $H_c(x)=P(Y_1\leq x, \delta=1)$, $y(t)=P(Y_1>t)$ and
\begin{equation*}
w(t)=\frac{1}{y(t)} \int{\frac{h_1(x)}{\widehat K(x-)}I(x>t)dH_c(x)},
\end{equation*}
where $h_1(y)=E(h(Y_1,Y_2)|Y_1=y).$ The proof of the next theorem  follows from Datta et al. (2010) with   choice of the kernel $h(Y_{1},Y_{2})=\min(Y_1, Y_2)$.
\begin{theorem}\label{thm5.1}
Assume  $E(\min^2(Y_1, Y_2))<\infty$, $\int {\frac{h_1(x)}{\widehat K^2(x-)}dH_c(x)}<\infty$ and\\  $\int_0^\infty w^2(t)\lambda_c(t)dt<\infty$.   As $ n \rightarrow \infty $, the  distribution of $\sqrt{n}({T}_{1c}-\mathcal{CRE})$ is Gaussian with mean zero and variance $4\sigma_{1c}^{2}$, where $\sigma_{1c}^2$ is given by
\begin{equation}
\sigma_{1c}^{2}=Var\Big(\frac{h_1(X)\delta_1}{\widehat K(Y-)}+\int w(t) dM_1^c(t)\Big).
\end{equation}
\end{theorem}

%

\section{Cumulative  extropy }\vspace{-.1in}
In this section, we discuss the estimation of negative cumulative extropy in uncensored and censored situations.  For a non-negative random variable $X$, negative cumulative  extropy is defined as (Tahmasebi and Toomaj,  2020)
\begin{equation}\label{cext}
   \mathcal{CE}=\frac{1}{2}\int_{0}^{\infty}(1-{F}^2(x))dx.
\end{equation}
\subsection{Uncensored case}
Let $X_1$ and $X_2$ be two independent  random variables with same distribution function $F$. Then the distribution function of    $\max(X_1,X_2)$ is  ${F}^2(x)$.  Hence we can represents $\mathcal{CE}$  as
\begin{equation}\label{adef}
  \mathcal{CE}=\frac{1}{2}E(\max(X_1,X_2)).
\end{equation}
A U-statistic given by
\begin{equation*}
  T_2=\frac{1}{n(n-1)}\sum_{i=1}^{n-1}\sum_{j=i+1}^{n}\max(X_i,X_j),
\end{equation*}is an unbiased and consistent estimator of $ \mathcal{CE}$. The above estimator can be written as
\begin{equation*}
  T_2=\frac{1}{n(n-1)}\sum_{i=1}^{n}(i-1)X_{(i)}.
\end{equation*}
\noindent Next we obtain the asymptotic distribution of $T_2$ and the proof the following result is similar to  Theorem 1.
\begin{theorem} \vspace{-0.1in}As $n\rightarrow \infty$,  $\sqrt{n}( T_2-\mathcal{CE})$  converges in distribution to  Gaussian random variable  with mean zero and variance $4{\sigma_2^2}$ where 
\begin{equation}\label{eq8} \vspace{-0.14in}
 \sigma_2^2 = V\Big(X{F}(X)+\int_{X}^{\infty}ydF(y)\Big).
\end{equation}
\end{theorem}

\subsection{Right censored case}
In this subsection, we find a simple estimator of $\mathcal{CE}$ in the presence of right censored observations. Using the same notations used in Section 2.2, a U-statistic for $\mathcal{CE}$ is given by
\begin{equation}\label{eq4.3}
 {T}_{2c}=\frac{1}{n(n-1)}\sum_{i=1}^{n}\sum_{j<i;j=1}^{n}\frac{\max(Y_i,Y_{j})\delta_i\delta_j}{\widehat{K}(Y_i-)\widehat{K}(Y_j-)}.
\end{equation}
\noindent Next results   establish  the consistency and asymptotic normality of $ {T}_{2c}$.
\begin{theorem}As $ n \rightarrow \infty $, $ {T}_{2c} $ converges in probability  to $\mathcal{CE}$.
\end{theorem}

\begin{theorem}\label{thm5.1}Let $h_1(y)=E(\max(Y_1,Y_2)|Y_1=y).$ Assume  $E(\max^2(Y_1,Y_2))<\infty$,\\ $\int {\frac{h_1(x)}{\widehat K^2(x-)}dH_c(x)}<\infty$ and $\int_0^\infty w^2(t)\lambda_c(t)dt<\infty$.   As $ n \rightarrow \infty $, the  distribution of $\sqrt{n}({T}_{c}-\mathcal{CE})$ is Gaussian with mean zero and variance $4\sigma_{2c}^{2}$, where $\sigma_{2c}^2$ is given by
\begin{equation}
\sigma_{2c}^{2}=Var\Big(\frac{h_1(X)\delta_1}{\widehat K(Y-)}+\int w(t) dM_1^c(t)\Big).
\end{equation}
\end{theorem}

%
\section{Empirical evidence}

To evaluate the finite sample performance of the proposed estimators we conduct a Monte Carlo simulation studies using R software. The simulation is repeated ten thousand times.
\newpage
\subsection{Cumulative Extropy}
\subsubsection{Complete data}
We generate observations from exponential, gamma, Weibull and  log normal  distributions. We find the bias and MSE of the proposed estimator based on  samples with size $n=10,20,30,40$ and $50$. We also compare the estimator with the estimator of $ \mathcal{CRE}$ proposed by Tahmasebi and Toomaj (2020) ($TT_1$). The results are given in Table 1.

\begin{table}[h]
	\caption{Bias and MSE  of  estimator of  $ \mathcal{CRE}$ }
	\scalebox{0.95}{
		\begin{tabular}{cccccccccccccc}\hline
\multirow{2}{*}{} & \multicolumn{4}{c}{Exponential (1)} & \multicolumn{4}{c}{Gamma (2,1)} \\ \hline
			\multirow{2}{*}{} & \multicolumn{2}{c}{$T_1$} & \multicolumn{2}{c}{$TT_1$}&\multicolumn{2}{c}{$T_1$}&\multicolumn{2}{c}{$TT_1$} \\ \hline
			$n$ &   Bias  & MSE&   Bias   & MSE  & Bias  & MSE&   Bias & MSE \\ \hline
			10 &0.00076&0.00879 & 0.02541&0.00776&0.00199&0.02647&0.19588&0.05776\\
			20& 0.00074&0.00416 &0.01283&0.00398&0.00168&0.01270&0.13943&0.03153\\
			30& 0.00038&0.00274&0.00861&0.00263&0.00124&0.00851&0.11270&0.02151\\
			40&  0.00018&0.00211&0.01283&0.00202&0.00094&0.00638&0.09819&0.01632\\
			50&  0.00004&0.00164&0.02541&0.00160&0.00036&0.00507&0.08221&0.01348\\ \hline
			\multirow{2}{*}{} & \multicolumn{4}{c}{Weibull (2,1)} &\multicolumn{4}{c}{Lognormal (0,1)} \\ \hline
\multirow{2}{*}{} & \multicolumn{2}{c}{$T_1$} & \multicolumn{2}{c}{$TT_1$}&\multicolumn{2}{c}{$T_1$}&\multicolumn{2}{c}{$TT_1$} \\ \hline
			$n$ &   Bias  & MSE&   Bias   & MSE  & Bias  & MSE&   Bias & MSE \\ \hline
			10 &0.00078&0.00431 & 0.12641&0.01951&  0.00115&0.01765&0.08175&0.02327\\
			20& 0.00058&0.00217 &0.09280&0.01105&  0.00098&0.00844&0.06710&0.01264\\
			30& 0.00054&0.00142&0.07621&0.00764&0.00049&0.00546&0.05867&0.00892\\
			40&  0.00017&0.00104&0.06703&0.00600&0.00030&0.00410&0.05367&0.00705\\
			50&  0.00001&0.00087&0.05987&0.00484&0.00024&0.00325&0.04867&0.00572\\ \hline
	\end{tabular}}
\end{table}

From Table 1, we can see that the bias and MSE of the proposed estimator is very low compared to that proposed by Tahmasebi and Toomaj (2020). Also bias and MSE decrease as sample size increases. This ensures the efficiency of the new  estimator.
\subsubsection{Censored data}
For generating  lifetimes we consider the same distributions used in uncensored case. In all the cases, the censored observations are generated from exponential distribution with parameter  $\lambda$ where $P(T>C)=0.2$. The  bias and MSE of the estimates when 20\% of the lifetimes are censored are given in Table 2. From Table 2, we observe that bias and MSE of the proposed estimator for censored observations are small and decreases with  sample size.
 \begin{table}[h]
 	\caption{Bias and MSE of  estimator of $ \mathcal{CRE}$ when 20\% of lifetimes are censored }
 	\scalebox{0.95}{
 		\begin{tabular}{cccccccccccccc}\hline
 			\multirow{2}{*}{} & \multicolumn{2}{c}{Exponential (1)} & \multicolumn{2}{c}{Gamma(2,1)}&\multicolumn{2}{c}{Weibull(2,1)}&\multicolumn{2}{c}{Lognormal(0,1)} \\ \hline
 			$n$ &   Bias  & MSE&   Bias   & MSE  & Bias  & MSE&   Bias & MSE \\ \hline
 			50 &0.08418&0.00830 & 0.17328&0.04083&0.08358&0.00914&0.26264&0.07059\\
 			75& 0.08287&0.00788 &0.17321&0.03822&0.08128&0.00852&0.26235&0.06979\\
 			100& 0.08222&0.00761&0.17318&0.03735&0.07956&0.00807&0.26228&0.06956\\
 			200& 0.08199&0.00741&0.17309&0.03585&0.07942&0.00762&0.26189&0.06895\\\hline
 			
 	\end{tabular}}
 \end{table}
 			

\subsection{Cumulative Extropy}
 The same simulation set up is employed to evaluate the efficiency of the estimator of negative cumulative  extropy. Here also we compare the estimator with the estimator of $ \mathcal{CE}$ proposed by Tahmasebi and Toomaj (2020) ($TT_2$). The  bias and MSE of the estimators  are given in Table 3. In this case also the proposed estimator has less bias and MSE compared to  the estimator of $ \mathcal{CE}$ proposed by Tahmasebi and Toomaj (2020).

 \begin{table}[h]
 	\caption{Bias and MSE of estimator of $ \mathcal{CE}$ }
 	\scalebox{0.95}{
 		\begin{tabular}{cccccccccccccc}\hline
 			\multirow{2}{*}{} & \multicolumn{4}{c}{Exponential (1)} & \multicolumn{4}{c}{Gamma (2,1)} \\ \hline
 			\multirow{2}{*}{} & \multicolumn{2}{c}{$T_2$} & \multicolumn{2}{c}{$TT_2$}&\multicolumn{2}{c}{$T_2$}&\multicolumn{2}{c}{$TT_2$} \\ \hline
 			$n$ &   Bias  & MSE&   Bias   & MSE  & Bias  & MSE&   Bias & MSE \\ \hline
 			10 &0.00118&0.01473 & 0.03877&0.01455&0.00074&0.02564&0.13489&0.04161\\
 			20& 0.00098&0.00722 &0.01938&0.00719&0.00052&0.01243&0.08878&0.02026\\
 			30& 0.00050&0.00491&0.00861&0.00488&0.00036&0.00827&0.06992&0.01344\\
 			40&  0.00038&0.00356&0.01283&0.00357&0.00032&0.00649&0.05864&0.01025\\
 			50&  0.00037&0.00297&0.02541&0.00295&0.00028&0.00493&0.05128&0.00783\\ \hline
			\multirow{2}{*}{} & \multicolumn{4}{c}{Weibull (2,1)} & \multicolumn{4}{c}{Lognormal (0,1)} \\ \hline
			\multirow{2}{*}{} & \multicolumn{2}{c}{$T_2$} & \multicolumn{2}{c}{$TT_2$}&\multicolumn{2}{c}{$T_2$}&\multicolumn{2}{c}{$TT_2$} \\ \hline
			$n$ &   Bias  & MSE&   Bias   & MSE  & Bias  & MSE&   Bias & MSE \\ \hline
			10 &0.00098&0.00845 & 0.15349&0.03316&0.00764&0.35391&0.16508&0.34391\\
			20& 0.00092&0.00419 &0.10558&0.01656&0.00430&0.18710&0.10203&0.18699\\
			30& 0.00031&0.00281&0.08592&0.01120&0.00385&0.12033&0.08764&0.12357\\
			40&  0.00013&0.00210&0.07303&0.00828&0.00064&0.08892&0.07814&0.09248\\
			50&  0.00001&0.00165&0.06441&0.00655&0.00044&0.07305&0.06620&0.07591\\ \hline
	\end{tabular}}
\end{table}

\subsubsection{Censored data}
The same parameter setup as in Section 4.1.2 is used in the simulation.  The bias and MSE of the proposed estimator of $ \mathcal{CE}$ when 20\% of lifetimes are censored are given in Table 4. From Table 4, we observe that the bias and MSE are little higher in case of exponential and gamma distributions. This may be due to the fact that exponential and gamma  distributions are positively skewed and  the estimator is based on maxima.

\begin{table}[h]
	\caption{Bias and MSE of estimator of $\mathcal{CE} $ when 20\% of lifetimes are censored}
	\scalebox{0.95}{
		\begin{tabular}{cccccccccccccc}\hline
			\multirow{2}{*}{} & \multicolumn{2}{c}{Exponential (1)} & \multicolumn{2}{c}{Gamma (2,2)}&\multicolumn{2}{c}{Weibull (2,1)}&\multicolumn{2}{c}{Lognormal (0,1)} \\ \hline
			$n$ &   Bias  & MSE&   Bias   & MSE  & Bias  & MSE&   Bias & MSE \\ \hline
			50 &0.19183&0.04122 & 0.28382&0.09566&0.11409&0.02219&0.00712&0.04865\\
			75& 0.19056&0.03923 &0.28137&0.08928&0.10925&0.01876&0.00639&0.03171\\
			100& 0.19048&0.03834&0.27948&0.08552&0.10690&0.01792&0.00447&0.02334\\
			200&  0.18831&0.03646&0.27659&0.08010&0.10540&0.01546&0.00227&0.01142\\ \hline
			
	\end{tabular}}
\end{table}

\section{Data analysis}

We apply the proposed estimation procedures to several real life data sets. The U-statistic based estimators for cumulative residual extropy and negative cumulative extropy are estimated in uncensored and censored cases.
\subsection{Complete data}
Three real life data sets are used for illustration.\\
\textbf{Example 1:} First, we consider the failure times of 84 mechanical components  which is reported in  Example 4.1 Tahmasebi and Toomaj (2020).  We obtain  $T_1=-0.9636$ and the corresponding estimator of Tahmasebi and Toomaj (2020) as ${TT_1} =-0.9494$. Cumulative extropy of this data is estimated as $T_2=1.9509$ and the estimate of the same by Tahmasebi and Toomaj (2020) as ${TT_2}=1.9251$.\\
\textbf{Example 2:} Next, we consider the data set on active repair time (in hours) for an airborne communication transceiver reported by Balakrishnan et al. (2009) and studied by Jahanshahi et al. (2020). The complete data set of 46 repair times is given in Jahanshahi et al. (2020) in Example 8. For this data estimator of cumulative residual extropy is calculated as $T_1=-0.7454 (TT_1= -0.6684)$ and the estimators of negative cumulative extropy is obtained as $T_2=2.8610({TT}_2=2.7380)$.\\
\textbf{Example 3:} Now we consider the data on the number
of million revolutions before failure for each of 23 ball bearings studied in Lawless (2011). The  data is given in Example 3.3.1 of Lawless (2011). For this data,  we find  $T_1=-25.6843 ({TT}_1=-17.1976)$ and $T_2=46.5365( {TT}_2=37.1432)$.

\subsection{Censored data}
Here also, we consider three real life data sets for demonstration.\\
\textbf{Example 1:} We consider the  data on survival times (in months) of patients with Hodgkin's disease given in Lawless (2011) (Example 3.2, Page 139) for illustration.  The data consists of 35 lifetimes out of which 9 are censored. Hence censoring percentage is around $25.7$. Now we estimate cumulative residual extropy as ${T}_{1c}=-3.4154$ and negative cumulative extropy as ${T}_{2c}=12.0849$ using the proposed procedures. \\
\textbf{Example 2:}  We examine the data on lifetimes of disk break pads on 40 cars studied in Lawless (2011) (see Table 6.11, Page 337). Out of the 40 observed lifetimes  22.5\% are  censored observations. In this case, cumulative residual extropy is estimated as ${T}_{1c}=-17.3063$ and negative cumulative extropy as ${T}_{2c}=25.3462$. \\
\textbf{Example 3:} We also analyse stanford heart transplant data available in R package `stanford2'. The data consist of 184 lifetimes where 72 (38.5\%) of them are censored. For this data, we obtain ${T}_{1c}=-711.5523$ and ${T}_{2c}=2028.036$.
\section{Further Discussions}
We gave alternative expressions for cumulative residual extropy and  negative cumulative  extropy. Using the alternative expressions we obtained estimators of these measures for complete and  censored cases.  We studied the asymptotic properties of the proposed estimators. Numerical illustrations are given  through Monte Carlo simulation study and real data analysis.

 Next, we give simple alternative expressions for different  extropy  measures. These alternative expression enables us to connect extropy  and reliability measures.   Using these alternative expressions one  can study the inference and other problems associated with these quantities.

Sathar and Nair (2021b) defined  dynamic survival extropy as
$$J_t(X)=-\frac{1}{2\bar{F}^2(t)}\int_{t}^{\infty}\bar{F}^2(x)dx.$$ For various properties of $J_t(X)$ interest readers  may refer Jahanshahi et. al.  (2020).
The  mean residual life function of $X$ (denoted by $m(t)$) can be express as
$$m(t)=E(X-t|X>t)=\frac{1}{\bar{F}(t)}\int_{t}^{\infty}\bar{F}(x)dx.$$ Hence using the survival function of  $\min(X_1,X_2)$, we can express $J_t(X)$ as
$$J_t(X)=-\frac{1}{2}E\left(\min(X_1,X_2)-t|\min(X_1,X_2)>t\right).$$
From the above expression, we can easily observe that $-2J_t(X)$ is the mean residual life function of  a series system having two identical components.

Kundu (2020) defined dynamic cumulative extropy given by
$$H_t(X)=-\frac{1}{2{F}^2(t)}\int_{0}^{t}{F}^2(x)dx.$$
The mean past life function of $X$ (denoted by $r(t)$) is given by
$$r(t)=E(t-X|X\le t).$$ Using some algebraic manipulations we can rewrite $r(t)$ as
$$r(t)=\frac{1}{{F}(t)}\int_{0}^{t}{F}(x)dx.$$ Hence using the distribution function of  function of  $\max(X_1,X_2)$, we can express $H_t(X)$ as
$$H_t(X)=-\frac{1}{2}E\left(t-\max(X_1,X_2)|\max(X_1,X_2)\le t\right).$$
Hence $-2H_t(X)$ is the mean past life function of a parallel  system having two identical components.

Weighted version of the  survival extropy is given by (see Sathar and Nair, 2021c)
$$J(X,w)=-\frac{1}{2}\int_{0}^{\infty}x\bar{F}^2(x)dx.$$
With simple algebraic manipulation we can express it as
$$J(X,w)=-\frac{1}{4}E(\min(X_1,X_2)^2).$$
The weighted version of the  cumulative extropy is given by
$$H(X,w)=-\frac{1}{2}\int_{0}^{\infty}x(1-{F}^2(x))dx.$$
Again with simple algebraic manipulation we obtain
$$H(X,w)=-\frac{1}{4}E(\max(X_1,X_2)^2).$$
Sathar and Nair (2021c) also defined weighted dynamic survival extropy as
$$J_t(X,w)=-\frac{1}{2\bar{F}^2(t)}\int_{t}^{\infty}x\bar{F}^2(x)dx.$$ Consider
\begin{eqnarray*}
  E(\min(X_1,X_2)-t^2|\min(X_1,X_2)>t) &=& \frac{1}{\bar{F}^2(t)}\int_{t}^{\infty}(y^2-t^2)2\bar{F}(y)dF(y)\\
   &=& \frac{1}{\bar{F}^2(t)}\int_{t}^{\infty}\left(\int_{t}^{y}2xdx\right)2\bar{F}(y)dF(y)\\
    &=& \frac{1}{\bar{F}^2(t)}\int_{t}^{\infty}\left(\int_{x}^{\infty}2\bar{F}(y)dF(y)\right)2xdx
    \\
    &=& \frac{2}{\bar{F}^2(t)}\int_{t}^{\infty}x\bar{F}^2(x)dx.
\end{eqnarray*}Hence
$$4J_t(X,w)=-E(\min(X_1,X_2)-t^2|\min(X_1,X_2)>t).$$
They also defined weighted dynamic cumulative extropy as
$$H_t(X,w)=-\frac{1}{2{F}^2(x)}\int_{0}^{t}x{F}^2(x)dx.$$Similar to above we obtain
$$4H_t(X,w)=-E(t^2-\max(X_1,X_2)|\max(X_1,X_2)\le t).$$
The alternative representations given above can be exploited for studying more about different extropy measures.



\end{document}